\pdfoutput=1
\documentclass[aps,prl,twocolumn,showpacs,superscriptaddress]{revtex4}
\usepackage{graphicx}
\usepackage{epstopdf}
\usepackage{latexsym}
\usepackage{amssymb}
\usepackage{amsmath}
\usepackage{amsfonts}
\usepackage{subfigure}
\usepackage{bm}
\usepackage{multirow}
\usepackage[normalem]{ulem}

\begin{document}

\title{Phase diagram and a possible unified description of 
intercalated iron
selenide superconductors }
\author{Yi-Zhuang You}
\affiliation{Institute for Advanced Study, Tsinghua University -Beijing, 100084, China}
\author{Fan Yang}
\affiliation{Department of Physics, Beijing Institute of Technology -
Beijing, 100081, China}
\author{Su-Peng Kou}
\affiliation{Department of Physics, Beijing Normal University -Beijing, 100875, China}
\author{Zheng-Yu Weng}
\affiliation{Institute for Advanced Study, Tsinghua University 
-Beijing, 100084, China}
\date{\today }

\begin{abstract}
We propose a theoretical description of the phase diagram and physical properties in A$_{2}$Fe$_{4}$Se$_{5}$-type (A=K, Tl) compounds based on a coexistent local moment and itinerant electron picture. Using neutron scattering and ARPES measurements to fix the general structure of  the local moment and itinerant Fermi pockets, we find a superconducting phase with s-wave pairing at the M pockets and an incipient sign-change  s-wave near the $\Gamma $ point, which is adjacent to the insulating phases. The uniform susceptibility and resistivity are found to be consistent with the experiment. The main distinction with iron pnictide superconductors is also discussed.
\end{abstract}

\pacs{74.20.Mn,71.27+a,75.20.Hr}
\maketitle

\emph{Introduction.---}The discovery of superconductivity in iron 
pnictides,\cite{kamihara} where the highest $T_{c}\simeq $ $55$ K\cite{ren} is about one third of a typical N\'{e}el temperature $T_{\mathrm{N}}\simeq 134$ K in the nearby magnetic phase,\cite{cruz} has renewed an intensive study of the interplay between superconductivity and antiferromagnetism.\cite{review} Most recently, a new class of iron-based superconductors, i.e., the intercalated iron selenides, has been synthesized,\cite{chen,fang} in  which a superconducting (SC) phase with $T_{c}\simeq 30$ K seems robustly present \emph{inside} an antiferromagnetic (AF) phase with $T_{\mathrm{N}}\sim 500$ K.\cite{muon,bao,xhchen} Such a coexistence with a large (one order of  magnitude) separation of the temperature scales, together with the presence of  an adjacent \emph{insulating} (instead of a metallic) phase with $T_{\mathrm{N}}$ essentially unchanged,\cite{bao,xhchen} make these materials distinctly different from the iron pnictides. It thus provides a unique opportunity to reexamine the possible SC mechanism underlying the iron-based superconductors.

\begin{figure}[htbp]
\centering
\subfigure[]{\includegraphics[height=0.11\textheight]{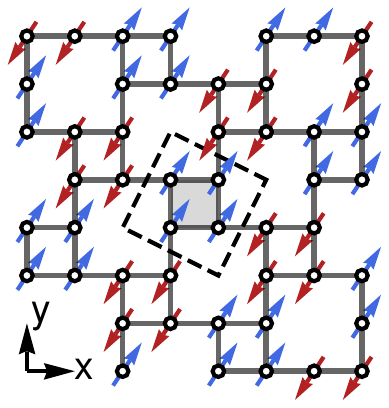}\label{fig:lattice}}\quad 
\subfigure[]{\includegraphics[height=0.11\textheight]{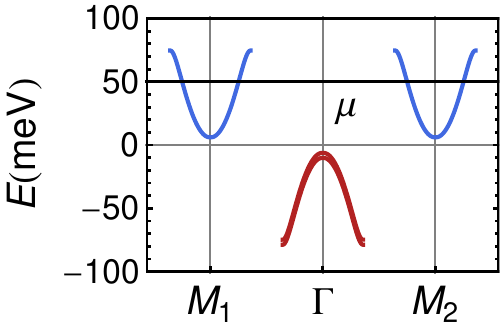}\label{fig:dispersion}}\newline
\subfigure[]{\includegraphics[height=0.12\textheight]{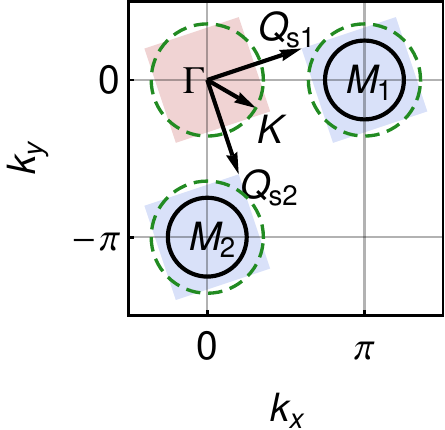}\label{fig:BZ}}\quad 
\subfigure[]{\includegraphics[height=0.12\textheight] {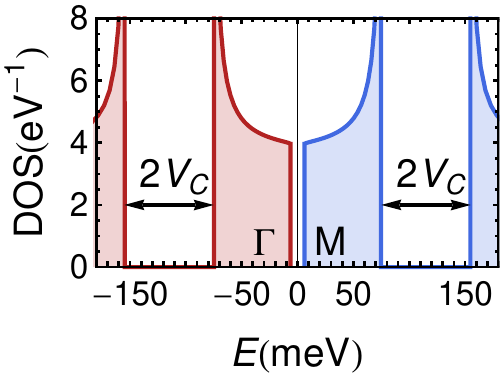}
\label{fig:DOS}}\newline
\caption{(Color online) (a) Top view of the Fe (denoted by circles) layer with 1/5 Fe-vacancies ordered in the right-handed rotation. The block AF ordering is indicated by the red/blue arrows. The shaded square is the 1-Fe unit-cell used through out this work. The dashed square is the enlarged $\sqrt{5}\times\sqrt{5}$ unit-cell for the vacancy ordering. (b) Bare band structure of the itinerant electrons. (c) $\Gamma $, $M_{1}$ and $M_{2}$ points in the 1-Fe BZ. Black circles around $M$ points indicate the Fermi surface at $\protect \mu =50$ meV. Shaded square regions are the folded BZ's for the right-handed lattice, while the dashed circles of a radius $K$ mark the ``averaged'' BZ's (see text). The AF wavevectors, $\mathbf{Q}_{s1}$ and $ \mathbf{Q} _{s2}$, are defined in the text. (d) The density of states  with the band gap induced by the Fe vacancy order.} \label{fig:model}
\end{figure}

In the intercalated iron selenides, e.g., A$_{2}$Fe$_{4}$Se$_{5}$ (A=K, Tl), the Fe atoms are basically arranged on a square lattice with 1/5 vacancy sites, which are ordered at $T_{\mathrm{S}}$, slightly higher than $T_{\mathrm{N}}$ where a block AF ordering occurs.\cite{bao} The vacancy orders into a $\sqrt{5}\times\sqrt{5}$ pattern\cite{fang,bao,vac1,vac2} with a chirality of either right-handed or left-handed [the former is shown in Fig.\,\ref{fig:lattice} with the block AF order illustrated as well]. The observed large magnetic moment ($\sim 3.3\, \mu_{\mathrm{B}}$ in K$_{2}$Fe$_{4}$Se$_{5}$\cite{bao}) suggests that the majority of the iron $3d$-electrons forms a local moment of $S\sim 2$, which is consistent with the LDA calculations\cite{dai,lu} where a large gap ($\sim 500$ \textrm{meV}) implies a Mott transition which stabilizes the large local moment.  The observation of spin-wave spectrum up to 220 meV \cite{PC Dai} further confirms the existence of the local moments.  On the other hand, the ARPES measurements\cite{ARPES1} have found the electron pockets at the $M$ points with an isotropic SC gap ($\sim 10$ \textrm{meV}), indicating the residual electron itineracy. The optical measurement  further indicates\cite{nlwang} a strong reduction of the itineracy in this  system as compared to the iron pnictides.

Based on these experimental facts, one may be tempted to treat\cite{Mott} the intercalated iron selenides as a doped AF/Mott insulator, which renders the iron-based superconductor a  multiband version of strongly correlated systems. However, there also exists a much simpler possibility for a multiband system with the Hund's rule coupling. Namely, via some kind of orbital-selective Mott transition, the majority of the $d$-electrons may form local moments with a large charge  gap, but the residual $d$-electrons may still remain quite itinerant at the Fermi  energy, which only perturbatively couple to the local moment rather than \emph{\ tightly} \emph{locking} with the latter as in a doped Mott insulator  case. Such a coexistent local moment and itinerant electron model has  been phenomenologically proposed\cite{KLW09,YYKW11} to  systematically describe the AF and SC states in the iron pnictides and achieved a consistent  account for the experiments.

 In this paper, by simply using the experimental input for the local  moment and itinerant electrons outlined above, we show that the  mechanism for both AF and SC states in A$_{2}$Fe$_{4}$Se$_{5}$ remains essentially the  same as in the iron pnictides by a coexistent model description. It predicts  an $s$-wave SC pairing at the $M$-pockets, while an incipient sign-changed $s$-wave pairing weakly induced around the $\Gamma $ point, even  though the hole pocket is below the Fermi energy on the electron doping side.  Here the pairing glue comes from mediating the spin fluctuations of the  local moments. The SC state generally persists in the metallic phase at low temperature, until at high or low doping where a competing charge-density-wave (CDW) or spin-density-wave (SDW) order sets in and turn the system into an insulator. It thus predicts a global phase diagram, whose low electron doping  regime is consistent with the experimental observations in A$_{2}$Fe$_{4}$Se$_{5}$. The corresponding uniform susceptibility and resistivity calculated  in this simple model are also in qualitative agreement with the  experiments. In the present approach, the essential distinction between the iron  pnictides and the intercalated iron selenides mainly lies in the (mis)match between the nesting momentum  of electron pockets and the characteristic momentum of local moment AF correlation.

\emph{Model.---}Our starting model Hamiltonian is of the same  general form as the one previously proposed for the iron pnictides:\cite {KLW09,YYKW11}
\begin{equation}
H=H_{\mathrm{it}}+H_{\mathrm{loc}}+H_{\mathrm{cp}}. 
\label{H}
\end{equation}
The first term $H_{\mathrm{it}}=\sum_{{\bm{k}}}\xi ({\bm{k}})c_ {{\bm{k}}}^{\dagger }c_{{\bm{k}}}$ describes the multiband itinerant  electrons created by $c^{\dagger }=(c_{\Gamma _{1}}^{\dagger },c_{\Gamma _{2}}^{\dagger },c_{M_{1}}^{\dagger },c_{M_{2}}^{\dagger })$, and $ \bm{k}$ is measured relative to the pocket center. The band structure $\xi ({\bm{k}})=\epsilon ({\bm{k}})-\mu $ is  phenomenologically written down based on the ARPES measurements:\cite{ARPES1} It  includes two degenerate hole-like bands around $\Gamma $ $(0,0)$ point and  two electron-like bands at $M_{1}$ $(\pi ,0)$ and $M_{2}$ $(0,\pi )$  points, respectively [with the nearest neighboring (nn) Fe-Fe lattice  constant taken as the unit], such that $\epsilon ({\bm{k}})$ will be a diagonal  matrix with diagonal elements as $(\epsilon _{\Gamma },\epsilon _{\Gamma }, \epsilon _{M},\epsilon _{M})$. We will stick to a particle-hole symmetric band structure $\epsilon _{\Gamma }= -\epsilon _{M}$  as shown in Fig.\,\ref{fig:dispersion} for the sake of simplicity, with  $ \epsilon _{M}({\bm{k}})={\bm{k}}^{2}/(2m)+\epsilon _{0}$, where  $m=6\, \mathrm{eV}^{-1}$ is the effective mass and $\epsilon _{0}=10\sim  15$ \textrm{meV} produces a small gap $2\epsilon _{0}>0$ separating $\Gamma $  and $M$ bands (note that $2\epsilon _{0}<0$ for the iron pnictide case\cite{KLW09,YYKW11}).

The Fe vacancy ordering will alter the above band  structure as the enlarged unit cell (cf. Fig.\,\ref{fig:lattice}) makes the Brillouin zone (BZ) folded to 1/5 of the original 1-Fe BZ, and opens up band gaps around the folded BZ boundaries. Considering two chiralities of the vacancy orders, the orientation of  a pocket BZ may be ``averaged'' to more isotropic as indicated by dashed circles in Fig. \ref{fig:BZ}, with an area of 1/10 of the 1-Fe BZ  characterized by a momentum $K=(2\pi /5)^{1/2}$. Such a band structure may be  fitted by $ \epsilon _{M}({\bm{k}})=\epsilon _{+}({\bm{k}})-\sqrt{\epsilon _{-} ({\bm{k}})^{2}+V_{C}^{2}}+\epsilon_{0}$, where $\epsilon _{\pm }({\bm{k}})= (|{\bm{k}}|^{2}\pm (2K-|{\bm{k}}|)^{2})/(4m)$ and $V_{C}$ controls the size of  the band gap ($V_{C}=V_{C0}=40$ \textrm{meV} at zero temperature). The corresponding density of states (DOS) is given in Fig.\,\ref {fig:DOS}, in which $\mu \sim 50$ \textrm{meV}  according to ARPES is still away from the edge of the band gap.

The second term in Eq.\,(\ref{H}) is $H_{\mathrm{loc}}=\sum_{ij}J_{ij} {\bm{M}} _{i}\cdot {\bm{M}}_{j}$, which generally describes the  superexchange interactions $J_{ij}$ between the iron local moments (denoted by $ {\bm{M}} _{i}$ at Fe site $i$). Here for A$_{2}$Fe$_{4}$Se$_{5}$, a block AF  order, instead of a ``stripe-like'' order  in the iron pnictides,\cite{cruz} has been identified by the neutron  scattering\cite {bao} as shown in Fig.\,\ref{fig:lattice}. Then ${\bm{M}}_{i}$ may be  ``coarse-grained'' within each enlarged unit-cell labeled by a position vector ${\bm{R}}$. Thus $ {\bm{M}}_{i} $ can be replaced by $\left( M/4\right) e^{i{\bm{Q}}_{s}\cdot {\bm {R}}}{ \bm{n}}({\bm{R}})$ where ${\bm{n}}({\bm{R}})$ is the unit vector for an effective spin $M\simeq 2\times 4=8$ in a block, with ${\bm {Q}}_{s}$ being either ${\bm{Q}}_{s1}=(3\pi /5,\pi /5)$ or ${\bm{Q}}_{s2}=(\pi /5,-3\pi /5)$ denoting the block AF wavevectors. Then the low-energy local moment fluctuations in $H_{\mathrm{loc}}$ may be properly  captured by a nonlinear $\sigma $-model in a Lagrangian form
\begin{equation}
\mathcal{L}_{\mathrm{loc}}=\frac{1}{2g_{0}}\left[ (\partial _{\tau } {\bm{n}} )^{2}+c^{2}(\nabla _{{\bm{R}}}{\bm{n}})^{2}+i\lambda ({\bm{n}}^ {2}-1)-\kappa ^{2}n_{z}^{2}\right]   \label{Ls}
\end{equation}
with $c$ as the spin wave velocity and $g_{0}$ the effective  coupling constant. In particular, $\kappa $ is an easy-axis anisotropy  parameter, which can effectively pin down the AF order at a finite $T_{\mathrm { N}}\sim 500$ \textrm{K}. The propagator for the ${\bm{n}}$ field is  given by \cite{KLW09,YYKW11} $D({\bm{q}},i\omega _{n})=-g_{0}/(\omega _ {n}^{2}+\Omega _{q}^{2})$ with $\Omega _{q}=$ $\sqrt{c^{2}q^{2}+\kappa ^{2}+ \eta ^{2}}$, in which $\eta ^{2}\equiv i\lambda ,$ determined by the condition $ \left \langle {\bm{n}}^{2}\right \rangle =1,$ vanishes at $T\leq T_{\mathrm{N}}$  where one finds $n_{0}\equiv \left \vert \left \langle {\bm{n}}\right \rangle  \right \vert $ quickly saturates to $1$ with the transverse spin fluctuations  gapped by $ \kappa $.

Finally, a local moment and itinerant electrons at each iron site  should be effectively coupled via a renormalized Hund's rule coupling $J_{H}$  in $H_{ \mathrm{cp}}=-J_{H}\sum_{i}{\bm{M}}_{i}\cdot {\bm{S}}_{i}$, where  ${\bm{S}} _{i}=\frac{1}{2}c_{i}^{\dagger }{\bm{\sigma}}c_{i}$ is the spin  operator for the itinerant electrons, and ${\bm{\sigma}}$ denotes the Pauli  matrices. Using the ``coarse-grained''  local moment, one finds $H_{\mathrm{cp}}=J_{0}\sum_{{\bm{k}},{\bm{q}},{\bm{P}}} {\bm{n}}_{ \bm{q}}\cdot c_{{\bm{k}}+{\bm{q}}\pm {\bm{P}}}^{\dagger }{\bm {s}}_{{\bm{P}} }c_{\bm{k}}$, where $J_{0}\propto J_{H}$, and ${\bm{P}}$ takes  either ${ \bm{Q}}_{s1}-(\pi,0)$ or ${\bm{Q}}_{s2}-(0,\pi)$ (with $M$ points as  the origin of momentum, cf. Fig. \ref {fig:BZ}). Here the spin-orbital matrices ${\bm{s}}_{\bm{P}}$ are given by
\begin{equation}
{\bm{s}}_{{\bm{P}}_{1}}=\left(
\begin{array}{cccc}
0 & 0 & {\bm{\sigma}} & 0 \\
0 & 0 & {\bm{\sigma}} & 0 \\
{\bm{\sigma}} & {\bm{\sigma}} & 0 & 0 \\
0 & 0 & 0 & 0
\end{array}
\right) \text{ \ }{\bm{s}}_{{\bm{P}}_{2}}=\left(
\begin{array}{cccc}
0 & 0 & 0 & {\bm{\sigma}} \\
0 & 0 & 0 & {\bm{\sigma}} \\
0 & 0 & 0 & 0 \\
{\bm{\sigma}} & {\bm{\sigma}} & 0 & 0
\end{array}
\right) .
\end{equation}

\begin{figure}[htbp]
\centering
\includegraphics[height=0.11\textheight]{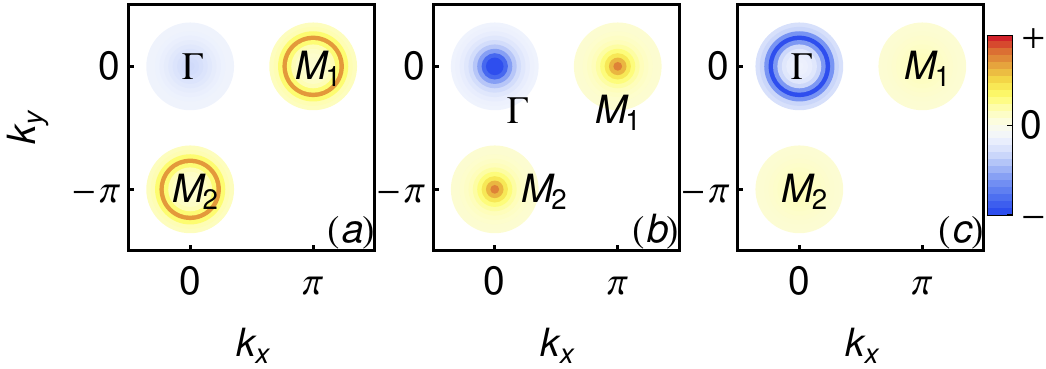}
\caption{(Color online) The pairing symmetry and strength characterized by $ f_{A}({\bm{k}})\Delta _{A}({\bm{k}})$: (a) The electron doped case at  $ \protect \mu =+50$meV (relevant to reality); (b) The undoped case at $\protect \mu  =0$meV; (c) The hole doped case at $\protect \mu =-50$meV.}
\label{fig:swave}
\end{figure}

\emph{Superconductivity.---} Similar to the previous consideration  for the iron pnictide case,\cite{YYKW11} the itinerant electrons will  experience an SC instability in the Cooper channel by exchanging the local  moment fluctuations. The effective pairing interaction is mediated by local  moment fluctuations $H_{\text{int}}=\frac{1}{2}\sum_{{\bm{k}},{\bm{k}}  ^{\prime }}c_{{\bm{k}}}^{\dagger }c_{-{\bm{k}}}^{\dagger }\Gamma  \left( { \bm{k}}-{\bm{k}}^{\prime }\right) c_{-{\bm{k}}^{\prime }}c_{{\bm{k}} ^{\prime }}$, with the vertex function given by $\Gamma ({\bm{q}})=J_{0}^ {2}\sum_{{ \bm{P}}}\text{Tr}D({\bm{q}}\pm {\bm{P}}){\bm{s}}_{{\bm{P}}}\otimes  { \bm{s}}_{{\bm{P}}}$. Here Tr stands for a summation over local  moment modes. $\Gamma ({\bm{q}})$ is a $64\times 64$ matrix determining  the pairing strength of the 64 modes, i.e., (2 spins$\times $4 pockets)$^{2}=64$.  To determine the pairing symmetry, we simply diagonalize $\Gamma  ({\bm{q}})$ and find the strongest attractive interaction in the channel  dominated by the \emph{spin-singlet intra-pocket} pairing, which involves 4  parameters: $ \Delta _{\Gamma _{1}}$, $\Delta _{\Gamma _{2}}$, $\Delta _{M_{1}}$,  $\Delta _{M_{2}}$ defined by $\Delta _{A}=(c_{{\bm{k}}A\uparrow }c_{-{\bm {k}} A\downarrow }-c_{{\bm{k}}A\downarrow }c_{-{\bm{k}}A\uparrow })/ \sqrt{2}$. Then according to the BCS theory, the linearized gap equation  reads $\Delta _{A}({\bm{k}})=\sum_{B,{\bm{k}}^{\prime }}\Gamma _{AB}\left( {\bm {k}}-{\bm{k} }^{\prime }\right) f_{B}\left( {\bm{k}}^{\prime }\right) \Delta _{B} \left( { \bm{k}}^{\prime }\right) $, where $A$, $B$ labels the pairing modes, and  $f_{\Gamma(M)}({ \bm{k}})=-(2\xi _{\Gamma(M)}({\bm{k}}))^{-1}\tanh (\beta \xi _{\Gamma(M)}({\bm{k}})/ 2)$ (where $\beta^{-1} \equiv k_{B}T$). Diagonalize the right-hand-side  of the gap equation, the greatest eigen value is found to be $2V_{\text{SC}}|f_ {\Gamma }f_{M}|^{1/2}$, with the corresponding eigen modes given by $ \Delta _{\Gamma _{1}}=\Delta _{\Gamma _{2}}\propto -|f_{\Gamma }|^{-1/2}$ and $ \Delta _{M_{1}}=\Delta _{M_{2}}\propto |f_{M}|^{-1/2}$, indicating \emph {$s$-wave} pairing with \emph{opposite sign} between $\Gamma $ and $M$  bands. Here $V_{ \text{SC}}=-J_{0}^{2}\langle D({\bm{k}}-{\bm{k}}^{\prime })\rangle _ {{\bm{k}} ,{\bm{k}}^{\prime }\in \mathrm{FS}}$ and $f_{\Gamma(M)}=\sum_{\bm{k}}f_ {\Gamma(M)}({\bm{k}})$. Figure\,\ref{fig:swave} shows the paring symmetry at  various dopings. The SC  is mainly $s$-wave on the Fermi surfaces around the $M$ points, but weak pairing order of opposite sign may still be induced in the hidden $\Gamma$ bands, reflecting essentially the same $s^{\pm}$-wave nature as in the iron pnictides.\cite {mazin}

\begin{figure}[htbp] 
\centering 
\subfigure[]{\includegraphics[height=0.12\textheight]{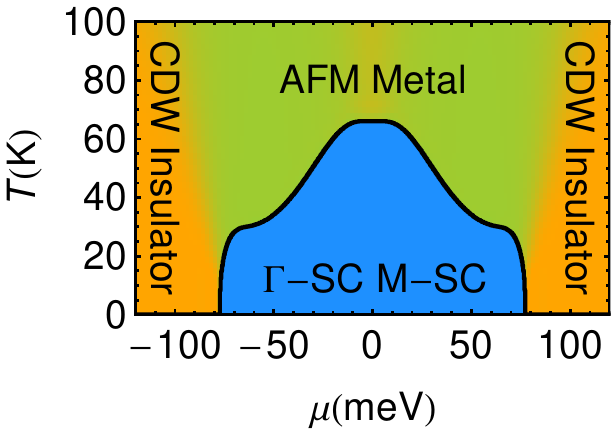} 
\label{fig:phase1}} \subfigure[]{\includegraphics[height=0.12
\textheight]{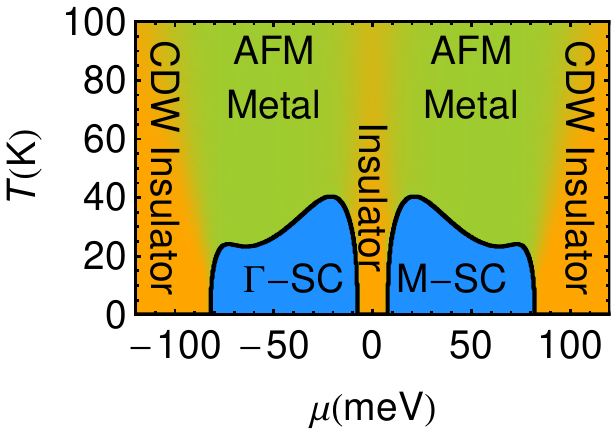}\label{fig:phase2}}
\caption{(Color online) The global phase diagram at different inter-pocket gaps: (a) $\protect \epsilon _{0}=10$ meV; (b) $\protect \epsilon _{0} =15$ meV. Notations, $\Gamma $-SC and $M$-SC, stand for the SC on $\Gamma $ pockets and $M$ pockets, respectively. The resistivity is calculated outside the SC phases, as shown by the color from yellow to green indicating the resistivity from high to low.} \end{figure}

The BCS mean field equation $2V_{\text{SC}}|f_{\Gamma }f_ {M}|^{1/2}=1$ (note that $f_{\Gamma(M)}$ are functions of $\mu$ and $T$) is solved numerically with fixed $V_{\mathrm{SC}}=0.36$ \textrm{eV}. Its solution trace out the boundary of the SC phase as shown in Figs.\,\ref{fig:phase1} and \ref {fig:phase2}, for $\epsilon _{0}=10$ \textrm{meV} and $\epsilon _{0} =15$ \textrm{meV}, respectively. In both cases, the SC phase eventually terminates when the vacancy-induced band edge is reached in the overdoped region, crossing over to an insulator caused by the Fe vacancy ordering.

The SC phase can extend into the small insulating region around $\mu=0$ (c.f. Fig.\,\ref{fig:phase1}), where the chemical potential rests in the band gap and the Cooper instability of the Fermi surface is not well-defined. It would be better to understand such a SC state as the condensation of the cooperons,\cite{Cooperon} formed by virtually exciting a pair of electrons from the valence band, and pairing them in the conduction band. If the energy cost to excite across the band gap can be compensated by the energy gain in the pairing, the cooperons will condense in the insulator. However, such a SC state is fragile and disappears (c.f. Fig.\,\ref{fig:phase2}) if the band gap $\epsilon_0$ is greater than a critical value $\epsilon _{0c}$, which can be seen from the following solution of the SC critical temperature $T_c$
\begin{equation} 
k_{\mathrm{B}}T_{c}=\frac{|\epsilon _{0}|}{2}\left[ \left( \frac{2W}{\epsilon _{0}}e^{-1/\lambda }-1\right) ^{2}-1\right] ^{1/2},  \label{eq:Tc}
\end{equation}
where $W$ is the typical band width of $\Gamma $ and/or $M$  pockets, and $ \lambda =2V_{\mathrm{SC}}(N_{\Gamma }N_{M})^{1/2}$ with $N_ {\Gamma(M)}$ the average DOS of the $\Gamma(M)$ bands, $N_ {\Gamma(M)}=m/(2\pi )$. $T_c$ will drop to zero at $\epsilon _{0c}=We^{-1/\lambda }$, which is of the same order as the zero-temperature SC gap $\Delta _{0}\simeq We^{-1/\lambda }$. Then it can be estimated that $\epsilon _{0c}\simeq \Delta _ {0}\sim 10 $ \textrm{meV}, according to the observed gap in the ARPES  experiment.\cite {ARPES1} An induced SC state due to the cooperon mechanism near $\mu=0$ provides a unique prediction for an explicit separation of local and itinerant electrons near the Fermi energy. 

At $\mu =0$, where the $\Gamma $ and  $M$ bands are both close to the Fermi energy, there is also a chance for  an incipient SDW order of the itinerant electrons to occur, as induced  by coupling to the block-AF-ordered local moments, albeit the  required momentum match between the two sub-systems is much weaker as compared  to the iron pnictide case.\cite{KLW09,YYKW11} In other words, the insulating state observed in A$_ {2}$Fe$ _{4}$Se$_{5}$-type compounds at low doping may well have a weak  SDW order of the itinerant electrons locking with the block AF order of the local  moment background.

\begin{figure}[t]
\centering
\subfigure[]{\includegraphics[height=0.10\textheight]{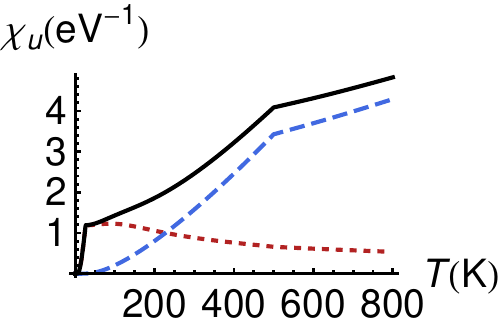}\label{fig:chi}}\quad
\subfigure[]{\includegraphics[height=0.12\textheight]{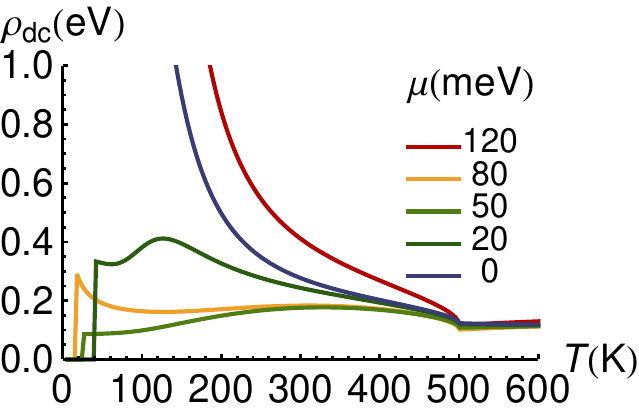} \label{fig:resistivity}}
\caption{(Color online) (a) The uniform magnetic susceptibility $  \chi_{u}= \chi _{\mathrm{it}}+ \chi _{\mathrm{loc}}$ at $\mu =50$ meV. The (red) dotted curve: itinerant electron part $ \chi _{\mathrm{it}}$; The (blue) dashed curve: local moment part $\chi _{\mathrm{loc}}$. (b) DC resistivity calculated at different $\mu$'s, corresponding to different electron dopings in Fig.\,\ref{fig:phase2}, including two insulating regimes  and the AF metal regime in between, with the SC transition at low temperatures.}
\end{figure}

\emph{Uniform susceptibility.---}The uniform magnetic  susceptibility composed of the contributions from both the itinerant electrons  and local moments: $\chi _{u}=\chi _{\mathrm{it}}+\chi _{\mathrm{loc}},$  similar to Ref. \cite{KLW09,YYKW11}, is shown in Fig.\,\ref{fig:chi} in  the metallic phase. Here $\chi _{\mathrm{it}}=-\sum_{{\bm{k}}}[n_{F}^ {\prime }(E_{\Gamma }({\bm{k}}))+n_{F}^{\prime }(E_{M}({\bm{k}}))]$ with  $E_{A}({ \bm{k}})=\sqrt{\xi _{A}({\bm{k}})^{2}+\Delta _{A}^{2}}$ is the  contribution from the itinerant electrons, which is suppressed by the s-wave  pairing in the SC state below $T_{c}$ (dotted curve). And local moments  contribute to: $ \chi _{\mathrm{loc}}=(\pi \beta c^{2})^{-1}[\Omega _{0}\beta (1-e^ {-\Omega _{0}\beta })^{-1}-\ln (e^{\Omega _{0}\beta }-1)]$ with $\Omega _{0} =\sqrt{ \kappa ^{2}+\eta ^{2}}$, which is qualitatively changed at $T_ {\mathrm{N} }=500$ \textrm{K} (dashed curve). The overall behavior of $\chi _{u}$ is in qualitative agreement with the experiments.\cite{xhchen,bao}

\emph{Resistivity.---}The resistivity for the electron doped case is calculated according to the following formula 
\begin{equation}
\rho _{\mathrm{dc}}^{-1}=\frac{\beta }{2}\sum_{\bm{k}}\frac{\bm {v}_{M}^{2}( \bm{k})}{\tau ^{-1}\left( \xi _{M}(\bm{k})\right) }\text{sech}^ {2}\frac{ \beta \xi _{M}(\bm{k})}{2}, 
\end{equation}
where $\bm{v}_{M}(\bm{k})=\partial _{\bm{k}}\xi _{M}(\bm{k})$  is the velocity of itinerant electrons in the $M$ bands, and the relaxation  rate is obtained from the self-energy through $\tau ^{-1}(\omega )=-\mathrm{Im} \Sigma (\omega )$, with $\Sigma (k)=-J_{0}^{2}\sum_{q}\text{Tr}D (q\pm\bm{P}){\bm{s}}_{\bm{P}}G(k+q){\bm{s}}_{\bm{P}}$. Here  $ G(k)=-\langle c_{k}c_{k}^{\dagger }\rangle $ stands for the itinerant electron propagator. Corresponding to the phase diagram shown in  Fig.\,\ref {fig:phase2}, the calculated resistivity is presented in Fig.\,\ref {fig:resistivity}. Here to simulate the charge ordering, we adopt a phenomenological model $V_{C}=V_{C0}\left[ 1-(T/T_{S})^{2}\right] ^{1/2}$ at $T<T_{\mathrm{S}}\simeq T_{\mathrm{N}}$. Again one  finds an overall qualitative agreement with the experimental measurements.\cite{xhchen,bao}  

\emph{Discussion.---}The discovery of iron-based superconductors,  especially the newly found intercalated iron selenides, has challenged the  notion that superconductivity generally competes with magnetism. Within the  BCS paradigm, an SC state coexisting and benefiting from magnetism is  only possible when they do not seriously compete for the electron  spectral weight near the Fermi energy. It was previously conjectured\cite {KLW09,OSMT1,OSMT2} that an orbital-selective Mott transition may take place among the  $3d$-electrons in iron-based superconductors such that the local  moment and the itinerant electron degrees of freedom are effectively separated, which can  eliminate the dynamic competition for the spectral weight at low energy,  while the long-wavelength fluctuation of the local moments provides with  the necessary pairing glue for the itinerant electrons. In the iron pnictide case,  due to a good momentum match (namely the AF wavevector well  connects the pockets at $\Gamma $ and $M$),  a joined AF/SDW ordering formed by both the local moment and itinerant electrons competes with the SC phase at low doping, and the SC phase gets suppressed in the magnetically ordered regime. In the present work, the SC phase can survive even in the presence of a static block AF order because the latter does not induce a strong SDW order due to the momentum mismatch (in fact, the $\Gamma $ pocket generally buries below the Fermi energy), such that the SC phase persists throughout the metallic regime coexisting with the magnetic ordering. Only at low doping or overdoping, the SC phase may get suppressed by insulating phases  possibly with a SDW order induced  by the local moment or a CDW order induced by the Fe vacancy ordering, which remain to be verified by future experiments. 

\begin{acknowledgments}
We would like to acknowledge stimulating discussions with W. Bao,  M.H. Fang, Z.Y. Lu, H. Ding, X. J. Zhou, H. Yao, and especially X.H. Chen. This work is supported by NSFC grant Nos. 10704008, 10834003 and 10874017, and the grants of National Program for Basic Research of MOST Nos. 2011CB921803, 2009CB929402 and 2010CB923003.
\end{acknowledgments}

\end{document}